\documentclass{elsart}
\journal{New Astronomy}

\usepackage{natbib}

\usepackage{graphics}
\usepackage{graphicx}
\usepackage{epsfig}

\usepackage{amssymb}

\def\lsim{\mathrel{\rlap{\lower4pt\hbox{\hskip1pt$\sim$}}
    \raise1pt\hbox{$<$}}}                
\def\gsim{\mathrel{\rlap{\lower4pt\hbox{\hskip1pt$\sim$}}
    \raise1pt\hbox{$>$}}}                

\begin{document}

\begin{frontmatter}



 \title{Searching for the Sources Responsible for Cosmic Reionization: 
Probing the Redshift Range 7$<$z$<$10 and Beyond}


\author{Daniel P. Stark$^1$ \& Richard S. Ellis$^1$}
\address{$^1$Dept. of Astronomy, California Institute of Technology, Pasadena CA 91125\\
E-mail: dps@astro.caltech.edu}

\begin{abstract}

We review recent observations that suggest that the global star
formation rate density of UV-bright galaxies is declining
monotonically with redshift over 3$<z<$7 and illustrate the
challenges that a continuation of this decline poses in explaining
the assembled stellar mass in several z$\sim$6 galaxies deduced
from recent Spitzer data. A plausible conclusion is a vigorous period of
yet earlier star formation. Prior to JWST and TMT, strong
lensing offers a unique probe of the extent of this earlier activity.
We discuss the first results of a blind spectroscopic survey of 10
lensing clusters for 8.5$<z<$10 Ly$\alpha$ emitters using
NIRSPEC on Keck.  We demonstrate this survey is achieving 
an (unlensed) sensitivity equivalent to a star formation rate of 
 0.1 $M_{\odot}$ yr$^{-1}$. A companion survey, now beginning 
 with HST and Spitzer, will target lensed $z$ and $J$-band 
 dropouts and probe a $\simeq$1 arcmin$^2$ region 1 magnitude deeper
than the UDF/NICMOS observations. In combination, both
surveys will provide the first constraints on the contribution
of early, low luminosity, sources to cosmic reionization.

\end{abstract}


\end{frontmatter}

\section{Introduction}

The reionization of neutral hydrogen which rendered the Universe 
transparent to UV photons and thereby terminated the so-called 
`dark ages' was a
landmark event in cosmic history. Determining the nature,
abundance and redshift distribution of the sources responsible for
this important transition represents the current frontier in
studies of the first galactic systems.

Recent developments strongly motivate a systematic search for
star-forming sources beyond z$\simeq$7. The WMAP
temperature-polarization cross correlation signal on large angular
scales \cite{Ko03} implies scattering of microwave photons
by free electrons from ionizing sources at $z\simeq$10-12
\cite{FK03}. The abundance of luminous UV-emitting $i-$band
\cite{Bu04}, $z-$band, and $J-$band \cite{B04b},\cite{B05} 
`drop-outs'
in the Hubble Ultra Deep Field at z$\simeq$6, 7 and 10 respectively, 
suggests a surprisingly continuous decline in the comoving density of 
emerging UV photons to z=10. The luminosity density of these UV-bright
sources seems insufficient to reionize the universe unless 
non-standard initial mass functions  \cite{SFP04b} or
very steep faint end-slopes for the luminosity function \cite{YW04} are
invoked. More likely is  the possibility of an intense earlier period of star formation.
Meanwhile, Spitzer observations have
uncovered a population of very massive (M$_{stellar}$ 
$\sim$ 10$^{10}$ - 10$^{11}$ M$_\odot$) galaxies with 
well-established stellar populations \cite{Eg05}, \cite{Ey05}, \cite{M05}, \cite{HY05}. 
Given the decline in the star formation rate density of UV-emitters,
the presence of this massive population requires explanation.  
While extinction and cosmic variance likely play a role in the 
apparent discrepancy, an earlier period of vigorous star formation 
may have contributed significantly to buildup these massive galaxies as 
well as producing a substantial fraction of the photons required for 
reionization.  

Finding this early population is one of the primary goals of observational 
cosmology.  With current facilities, traditional 
search methods are poorly equipped to search beyond z$\simeq$7.
Even with HST and 8-10 meter class telescopes, continuum (`drop-out') 
and narrow-band 
imaging Lyman $\alpha$ searches only probe the most luminous, rare, sources 
at these redshifts; the contribution to the reionization flux from these
sources may well be small. The study of lower luminosity 
sources rendered visible by the strong
gravitational lensing induced by clusters with well-constrained
mass models can greatly assist in this work. A typical lensing cluster 
magnifies sources by $\times$5-10 over 1 arcmin$^2$ for sources at $z>$7. 
Moreover, on the  critical line,
faint signals are boosted $>$$\times$30 and spectroscopy and more
detailed studies are often feasible at otherwise impossible (unlensed)
limits. For several objects recently located via this technique 
\cite{El01}, \cite{Sa04}, \cite{Kn04},
stellar continuum slopes, Ly$\alpha$ profiles and star formation
rates have been determined for sources whose intrinsic flux is close 
to the UDF limit, $I_{AB}\simeq$30. Perhaps most remarkably,
such sources have been magnified so as to become accessible
by Spitzer \cite{Eg05} yielding luminosity-weighted ages.

Until the era of TMT/JWST, scanning the critical lines of massive
clusters provides possibly the only reliable means of estimating the 
abundance of low luminosity star forming sources in the 7$<z<$10
interval with minimal contamination.  In these proceedings, we
discuss two such surveys: in $\S$4, we report on the progress of
 a Keck spectroscopic survey along cluster critical lines for Ly$\alpha$
emitters at z$>$8, and in $\S$5 we present a new 110 orbit Hubble 
imaging survey for lensed low-luminosity galaxies at z$\sim$7-10. 

\section{Evidence for a Declining Star Formation Rate}

Optical and near-infrared imaging with the Hubble Space Telescope
 have tentatively provided the first limits on the visible star 
formation activity in the era 6 $\lsim$ z $\lsim$ 10. Galaxies 
are identified at these redshifts 
by extending the Lyman-break technique \cite{SPH95}, \cite{S96} to 
z$>$6, and their star 
formation rates are estimated by measuring their rest-frame 
ultraviolet emission, which is dominated by young massive stars.

The abundance of z$\sim$6 galaxies (identified as $i$'-dropouts) 
has been calculated in the 400 arcmin$^2$ 
GOODS-North and GOODS-South fields down to z$_{AB}$$<$25.6 
\cite{St03},\cite{St04}, in ACS parallel observations covering 10 
arcmin$^2$ down to z$_{AB}$$<$28.3 \cite{YWC03}, in ACS 
GTO fields spanning 46 arcmin$^2$ reaching down to 
z$_{AB}$$<$27.3 \cite{B04a}, in the ACS parallels to the 
NICMOS UDF observations which span 21 arcmin$^2$ down 
to z$_{AB}$$<$28.1, and in the 11.3 arcmin$^2$ UDF down to  
z$_{AB}$ $<$ 28.5 \cite{Bu04} and down to z$_{AB}$ $<$ 29.2
\cite{YW04}.  In addition, the exceptional depth of the HST NICMOS images 
of the UDF allows galaxies at z$\sim$7 and z$\sim$10 to be identified  
as z-drops and J-drops, respectively \cite{B04b}, \cite{B05}.

The general consensus emerging from these studies is that the apparent 
star formation rate density, uncorrected for extinction (integrated 
down to L$_{UV}$$>$0.1 L$^\star_{z=3}$) is $\sim$3-6 times less 
at z$\sim$6 than at z$\sim$3 and continues to decline out to 
z$\sim$10 (integrated down to L$_{UV}$$>
$0.3L$^\star_{z=3}$) (Fig 1). The declining star formation rate poses a 
challenge for models suggesting star-forming galaxies at z$\sim$6 
reionized the Universe. The low-abundance of high-z dropouts can 
be reconciled with such models if the galaxies have low metallicity
or a non-standard initial mass function \cite{SFP04b},
 or if the UV luminosity function has a very steep faint-end slope 
\cite{YW04}, such that the bulk of ionizing photons 
are released by intrinsically faint galaxies that lie below 
the 5$\sigma$ sensitivity limit of the UDF.  Alternatively, the 
bulk of reionization may occur at yet earlier times, requiring the 
presence of a significant abundance of objects at z$\gg$6, or 
conceivably reionization was not accomplished by young star-forming
galaxies.

\begin{figure}[!t]
\centerline{\psfig{file=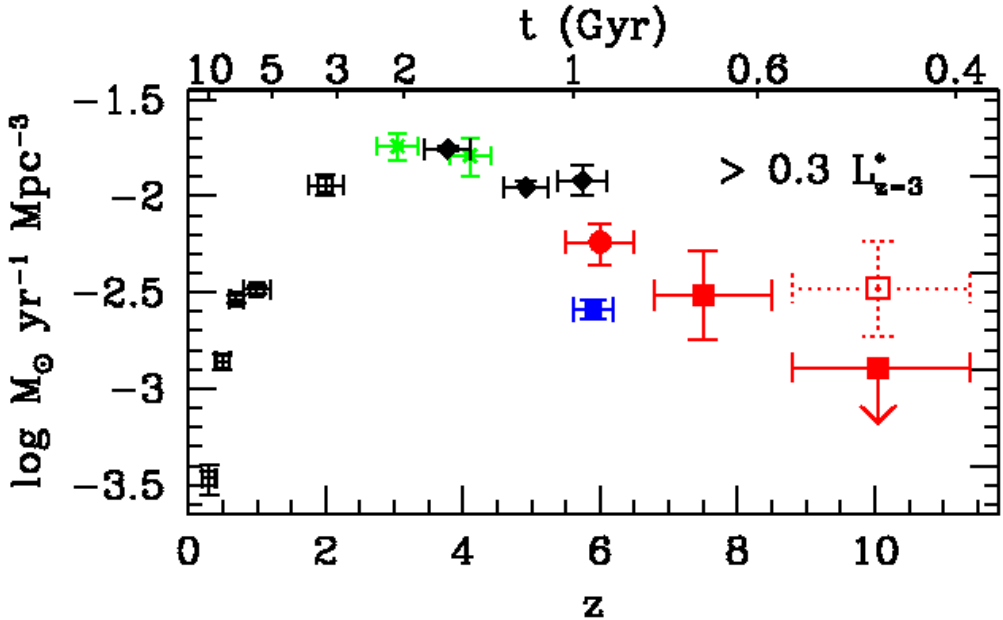,width=5.4in}}
\caption{ The co-moving volume density of
luminous UV emitting galaxies located via the `drop out' technique with
recent UDF estimates from \cite{Bu04} and \cite{B04b}, \cite{B05}, uncorrected for extinction, as 
presented in \cite{B05}. The integrated abundance to $z\simeq$10, while
uncertain, appears to be insuffient for cosmic reionization. Possible
explanations include an abundant population of lower luminosity sources 
or vagaries of cosmic variance in the small UDF field. Our surveys 
($\S$4 and $\S$5) test both hypotheses.
}
\label{fig:z7sed}
\end{figure}

\section{The Crucial Role of Spitzer}

Detecting light from the most distant galaxies known in the universe 
has been one of the most remarkable and important feats of the IRAC
instrument onboard the 85 cm Spitzer Space Telescope.  The IRAC filters
at 3.6-8.0 $\mu$m probe the rest-frame optical and near-infrared at 
z$>$6, providing the first glimpse of light from established 
stellar populations of galaxies at such high redshift. Combining 
these data with deep broadband optical photometry from HST and 8-10 
meter class ground based telescopes, spectral energy distributions 
(SEDs) can be constructed and compared with population synthesis models 
to constrain the age, star formation history, and stellar mass of 
galaxies at z$\sim$6-7.  In providing the opportunity 
to characterize the stellar mass density at 
z$\sim$6-7, Spitzer allows the first constraints to be placed 
on the star formation rate density integrated over the first 
800 Myr of the Universe. 

The first demonstration  of an established stellar population at high
redshift was via the strongly-lensed z$\sim$6.8 galaxy in Abell 2218 
\cite{Eg05}.
 The initial identification of this triply-lensed source was 
based on detections in the F814W, F850LP, and F160W HST 
filters (0.801 $\mu$m, 0.905 $\mu$m, and 1.612 $\mu$m 
respectively). Subsequent detections were made at 3.6 $\mu$m 
and 4.5 $\mu$m with Spitzer for one of the components of the lensed system 
and at 1.1 $\mu$m with HST for two of the three lensed images.   With the new 
photometric data, the galaxy's SED was constructed and compared to 
model SEDs generated with the GALAXEV population synthesis code
\cite{BC03}.  Best-fit SEDs of this source (Fig 2) 
are obtained by varying the galaxy age and redshift for a 
given star formation history and minimizing $\chi$$^2$.  The
results suggest that the age of the galaxy is between 40-450 Myr,
corresponding to a formation redshift of z$_f$ $\sim$ 7 - 12.  In 
each case the best-fit age of the galaxy is greater than the
assumed e-folding star formation timescale, indicating that
the galaxy would have been more luminous in the ultraviolet at 
earlier times during its active star-forming phase. Given the 
small survey area required for its discovery, such sources
may have a reasonably high surface density suggesting 
that a population 
of intense star-forming galaxies (10-100 M$_\odot$ yr$^{-1}$) 
may be very common at z$>$7, contributing significantly to 
the UV photon budget required for reionization.  

\begin{figure}[!t]
\centerline{\psfig{file=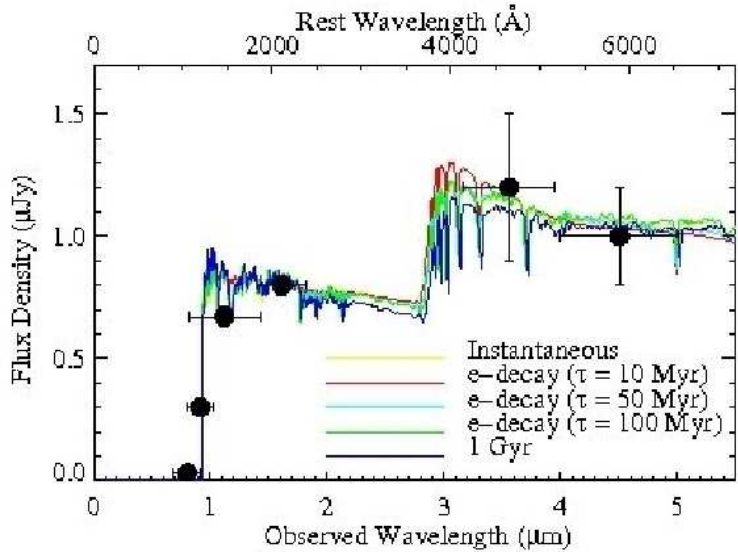,width=4.4in}}
\caption{  SED of the $z\simeq$6.8 lensed 
source in Figure 2 from \cite{Eg05}. The combination of Spitzer and
HST photometry spans the rest-frame UV to optical determining the
luminosity-weighted stellar age ($\simeq$100 Myr) suggesting a
credible source of UV photons since $z\simeq$10. 
}
\label{fig:z7sed}
\end{figure}
 
The key feature which makes the lensed source in Abell 2218 of 
great diagnostic value is the Balmer break observed between the NICMOS F160W 
and 3.6 $\mu$m bands; this indicates an age $\gsim$ 100 Myr and a declining
star formation rate.  Similar signatures have been subsequently seen in 
more massive systems at z$\sim$6-7 (M$_{stellar}$ $\simeq$ 10$^{10}$ 
M$_\odot$).  The SEDs of four $i$-drops in GOODS-South, 
spectroscopically confirmed by us at Keck, are constructed in \cite{Ey05}. 
Two of the four galaxies have robust Spitzer 
detections at 3.6 $\mu$m and 4.5 $\mu$m.  Comparisons to population
synthesis models \cite{BC03} yield stellar mass estimates
of 2-4 $\times$ 10$^{10}$ M$_\odot$ and ages between 250-650 Myr 
(z$_f$ $\sim$ 7.5-13.5).  Very recently, similarly massive systems
have been presented for a larger sample in the UDF \cite{HY05}.
In order to build up such a stellar mass, the average SFR at earlier times 
must be greater than that observed at $z\simeq$6.

Perhaps the most intriguing high-redshift object studied 
with Spitzer is a z$\sim$7 galaxy in the Hubble UDF with a 
stellar mass claimed to exceed 10$^{11}$ M$_\odot$ \cite{M05}. 
The high-redshift interpretation 
is favored as a result of the strong break in flux between the K-band 
and 3.6 $\mu$m  (interpreted as the rest-frame 3646 $\AA$ Balmer break) 
and the null detection in deepest optical broadband images 
ever obtained (3$\sigma$ detection limit of 31.5 in AB mags over 
a 1.5 arcsec aperture in the combined BViz ACS images of the UDF).   
The best-fitting population synthesis models have very little dust 
extinction and ages of several hundred Myr, placing the 
epoch of formation of the bulk of this galaxy's stellar mass
at z$_f$ $>$ 10. Even if star formation occured in several sub-galactic
units which subsequently merged, remarkably high star formation 
rates ($>$ 1000 M$_\odot$ yr$^{-1}$) may be required to assemble the 
stellar mass of this system by z$\sim$7.  

\section{A Systematic Survey for Strongly-Lensed Ly$\alpha$ emitters at z$>$8}
\label{sec:nspec}
\subsection{Survey Strategy}
The above discussion provides a clear motivation for searching 
for earlier activity and, prior to JWST and TMT, lensing offers 
a reasonable prospect of making some progress.  We have undertaken 
a Keck spectroscopic survey for strongly-lensed star-forming 
sources at z$>$8.  Using NIRSPEC in the J-band (sensitive 
to Ly$\alpha$ emitters at 8.5$<$z$<$10.2), we scan the critical 
lines of $\sim$10 lensing clusters.  By 
only observing along regions of very high magnification ($\times$30
across most of the slit), we become sensitive to feeble sources 
producing stars with rates as low as $\sim$0.1 M$_\odot$ yr$^{-1}$. 
All clusters used in our survey have spectroscopically-confirmed 
redshifts for a number of multiple images and arcs, allowing for the
derivation of detailed magnification maps for high-redshift sources 
(Figure 3).    

\begin{figure}[!t]
\centerline{\psfig{file=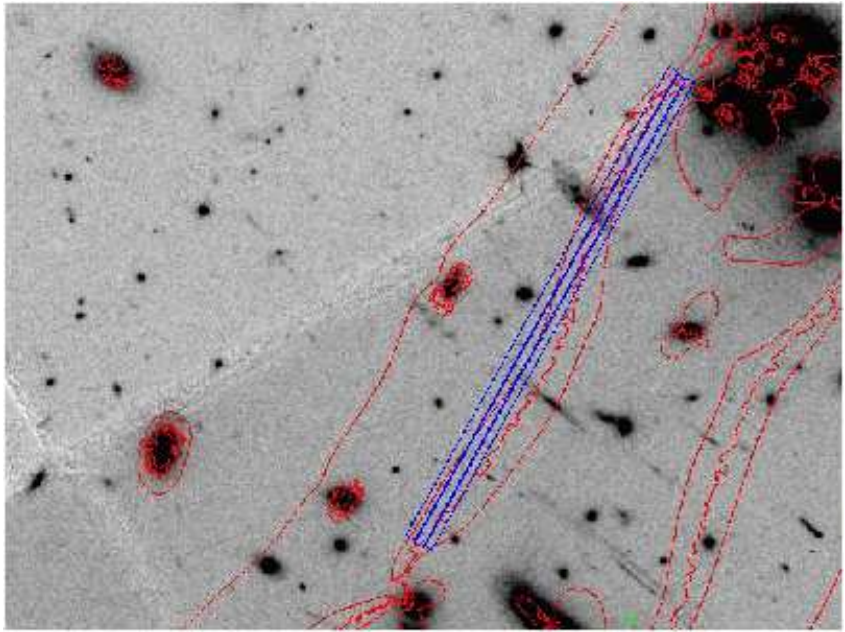,width=4.4in}}
\caption{The Keck NIRSPEC blind survey: slit positions observed 
toward Abell 68.  Amplification 
contours ($\times$5,30) are overlayed on the HST WFPC2 R-band
image.  The magnification appropriate for z$>$6 background sources
is $>$30 for the entire extent of each of the 
three slit positions. 
}
\label{fig:amp}
\end{figure}

Redshift identification of candidates is aided by several factors. 
The cluster lensing model constrains the redshift of any  
strongly-lensed source for which multiple images are detected 
via the observed image flux ratios and positions.    
In addition, by using NIRSPEC to search for Ly$\alpha$ emitters 
in the same clusters studied with LRIS on Keck in the optical 
\cite{Sa04}, we are able to correlate emission 
lines seen in the $J$ band with existing optical spectroscopic data to 
clarify possible interlopers which are not multiply-imaged.  
Further, deep HST optical broadband imaging is available for all 
of our clusters.  Ly$\alpha$ emitters at z$>$8 should be virtually
invisible in the optical due to IGM absorption.
Hence, the presence of optical flux spatially coincident with a 
candidate z$>$8 Ly$\alpha$ emitter would indicate that the candidate
is at low-redshift.  

\subsection{Results and Discussion}

Twelve Keck nights have been allocated for this project.  Nine nights have 
been completed thus far, of which 5.5 have been clear. During
these 5.5 nights, 31 slit positions were mapped 
toward seven clusters with an average integration time of 1.4 hours each. 
Each slit position consists of 6 to 9 ten minute exposures that 
were dithered parallel to the length of the slit.

Spectra are flat-fielded and sky-subtracted with IDL routines 
following the optimal spectroscopic reduction techniques presented in 
\cite{K03}.  We determine the pointing of each exposure on 
the sky by registering images from the NIRSPEC slit-viewing camera 
to HST images of the same field. 
Offsets between exposures of a given slit position are 
calculated from the registered slit viewing camera images, 
and the spectra are subsequently shifted and combined.  In 
Figure 4, we present the reduced final two-dimensional spectra of three 
slit position toward Abell 68.  

\begin{figure}[!t]
\centerline{\psfig{file=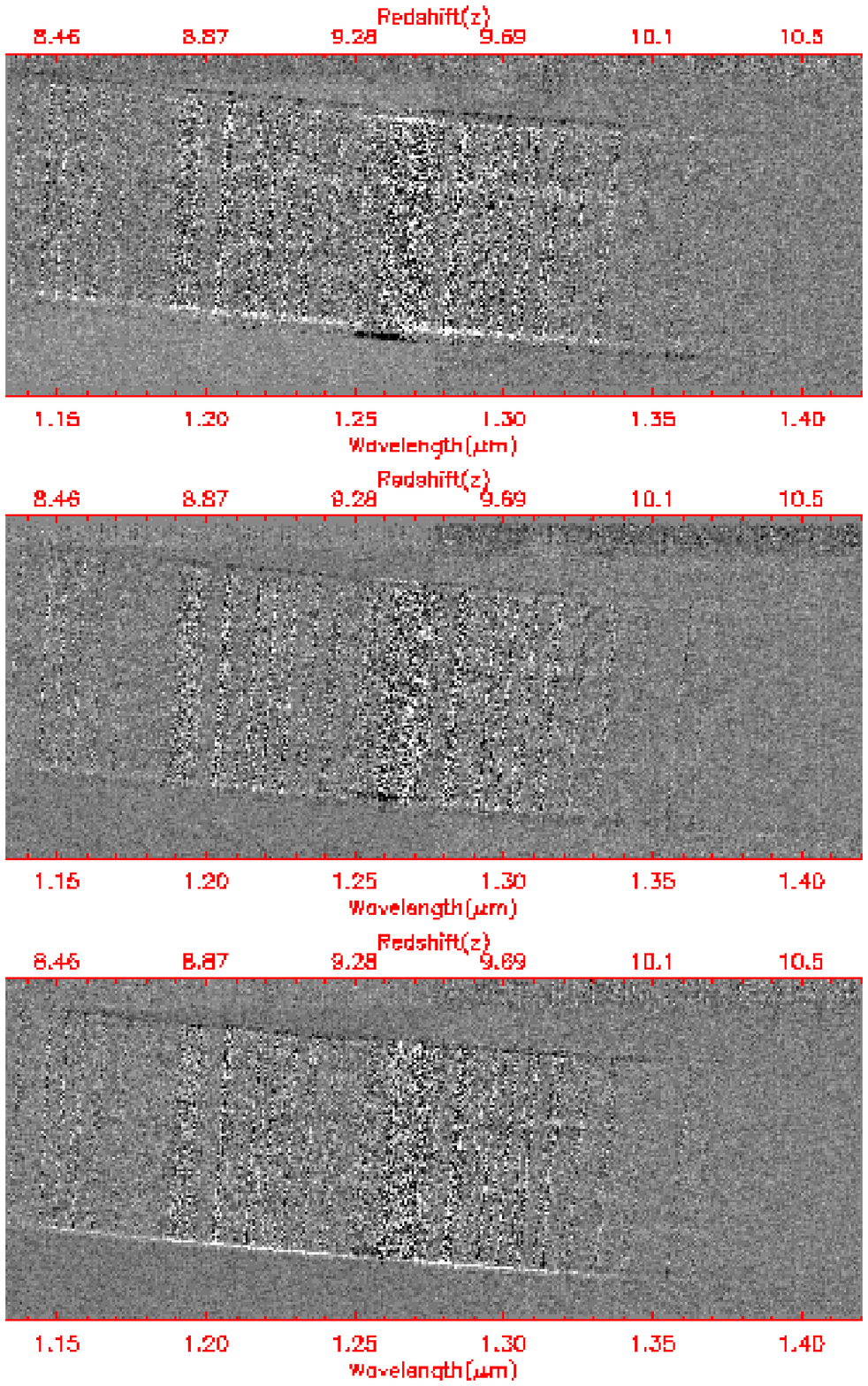,width=2.5in}\psfig{file=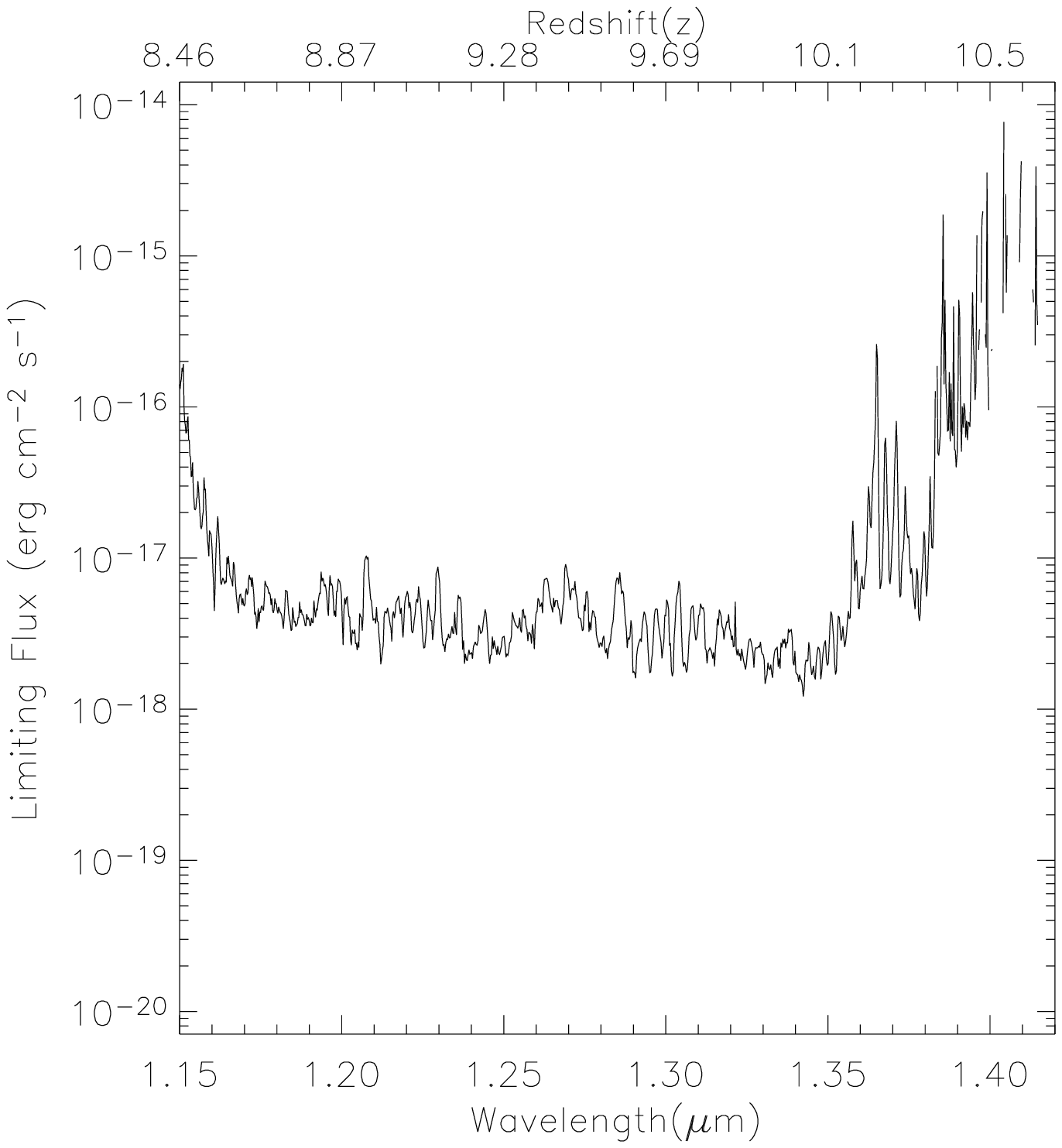,width=3in} }
\caption{Left: Reduced two-dimensional spectra of three slit positions
toward Abell 68. A faint continuum trace from a foreground spiral galaxy 
is apparent in each of the three slit positions.  Right:  The minimum detectable
flux (5$\sigma$) for an observed Ly$\alpha$ line as a function of
wavelength (and redshift) from a 1.5 hour exposure with NIRSPEC.
Depending on the local magnification, this observed limit of
$<$10$^{-17}$ erg cm$^{-2}$ s$^{-1}$ corresponds to a star formation 
rate of $\simeq$ 0.03-0.1 $M_{\odot}$ yr$^{-1}$. 
}
\label{fig:spec}
\end{figure}

The sky-subtracted two-dimensional spectra were independently 
inspected for candidates by the authors.  
A catalog of fifteen possible high-z Ly$\alpha$ emitters was constructed
and cross-checked
with deep HST optical and ground-based near-IR images for corresponding
broadband emission.  Most candidates lack convincing emission
in the ground-based near-IR broadband images.  Final confirmation 
of these sources awaits deeper HST near-IR imaging (see $\S$ 5).  

Our limiting Ly$\alpha$ line flux is defined to be the 
signal in an aperture 1.0 arcsec by 12 $\AA$ that is five times
the root-mean-square fluctuation in an aperture of that size.  The spatial 
dimension was chosen to be roughly twice the typical seeing disk 
of the observations, and the spectral dimension was chosen to match
the typical Ly$\alpha$ line width of high-redshift galaxies, $\sim$
300 km s$^{-1}$ \cite{Sa04}.  Typical 5$\sigma$ limiting line 
fluxes between OH lines 
range between $f_{\alpha}$=3 $\times$ 10$^{-18}$ erg cm$^{-2}$ s$^{-1}$
and 1$\times$ 10$^{-17}$ erg cm$^{-2}$ s$^{-1}$ (Figure 4). The corresponding 
characteristic limiting Ly$\alpha$ line luminosity of our survey is 
$\sim$10$^{41}$ erg s$^{-1}$ after correction for the typical $\times$30 
magnification.  Converting this line luminosity to a star formation
rate, assuming 1 M$_\odot$ yr$^{-1}$ of star formation produces
 10$^{42}$ erg s$^{-1}$ of luminosity in the Ly$\alpha$ line 
\cite{K98}, we find the typical unlensed limiting star 
formation rate is $\sim$0.1 M$_\odot$ yr$^{-1}$. 

To obtain a rough estimate of the surface density of galaxies we 
would expect to find in our survey if low-luminosity galaxies dominate the 
reionization process, we consider two scenarios: 1) 
an extended reionization period (z=20 to z=7, $\Delta$t=575 Myr) 
with a moderate ionizing 
photon escape fraction (f$_c$=0.5) and 2) a relatively quick 
reionization period (z=10 to z=7, $\Delta$t=300 Myr) with a lower 
ionizing photon escape fraction (f$_c$=0.1).  With these assumptions, 
we find that the surface density of sources in our survey must exceed 
200 arcmin$^{-2}$ for the extended reionization case and 2000 arcmin$^{-2}$
for the abrupt reionization scenario (Figure 5).  Given our expected 
survey area of 0.015 arcmin$^2$ (unlensed), upon completion of our survey 
we should detect several sources under a range of reionization scenarios.  

We note that realistically, galaxies of all luminosities contribute to 
the reionization 
process.  If the number density of galaxies continues to rise steeply toward
 star formation rates below the limiting rate of our survey, the number density 
of star-forming sources required 
for reionization is lower than we estimate above.  We consider this possiblity 
in more detail in a future paper (Stark et al. 2005, in prep).

At the time of writing about half of our survey data has been analyzed
suggesting that the abundance of galaxies at these early times is not 
significantly greater than that required for reionization under 
the assumptions mentioned above; however, it is
clear from Figure 5 that our ability to detect reionization sources and
provide meaningful constraints on the contribution of intrinsically faint 
galaxies to reionization at z$>$8 will be significantly enhanced upon 
completion of our survey.  

Figure 5 also provides an opportunity for us to consider the challenge
of explaining the surface density of massive stellar systems being found 
by Spitzer at $z\sim$6  \cite{Eg05}, \cite{Ey05}, \cite{HY05}. For each of the
surveys (marked), we estimate the lower limit to the mass density from
the sources so far seen and infer an equivalent lower limit to the (constant) 
star formation rate density required over the interval 6$<$z$<$20 to account 
for this mass. Not only are the estimates close to being adequate for 
reionization but, if the decline in Fig. 1 is correct, the presence of a few
massive galaxies in the small fields so far surveyed already suggests 
the need for a very short intense period of activity at high redshift.

\begin{figure}[!t]
\centerline{\psfig{file=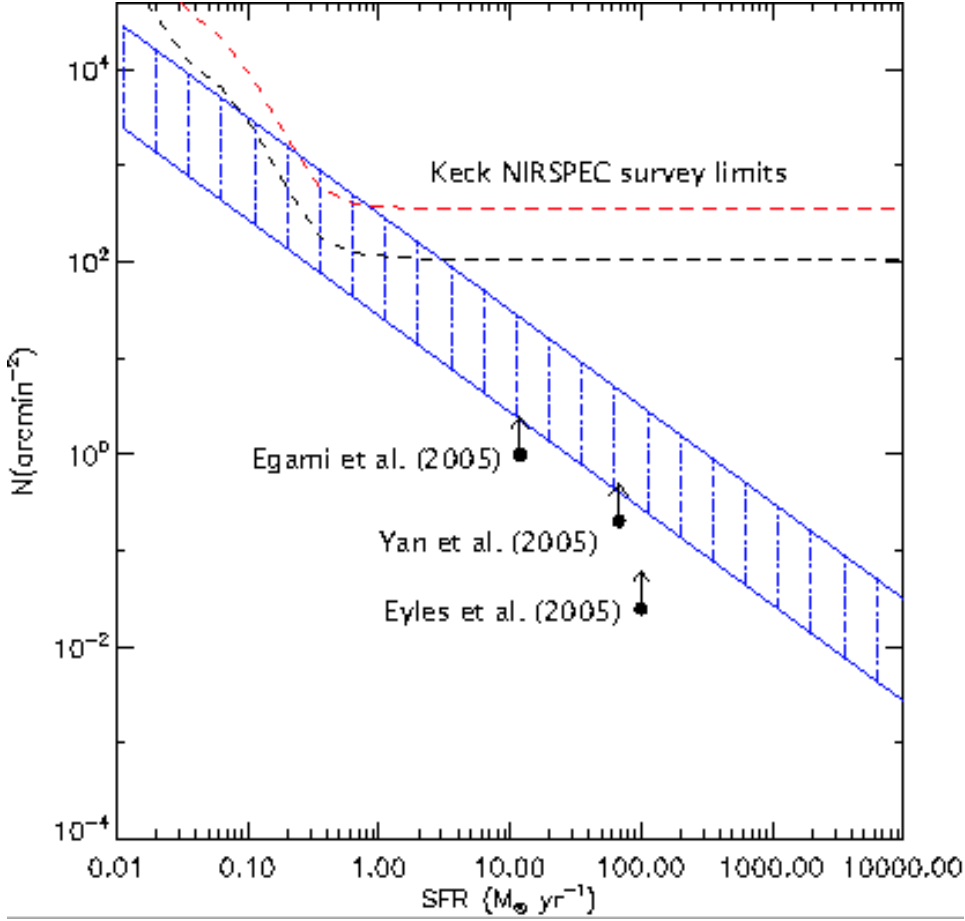,width=5in}}
\caption{Estimates of the surface
density of sources necessary for cosmic reionization over 7$<z<$10. 
The blue hatched area represents the likely range for various scenarios;
the lower bound represents an extended phase (7$<z<$20) and 
moderate ionizing photon escape fraction (f$_c$=0.5) and the upper 
bound represents a rapid phase (7$<z<$10) and a low ionizing photon 
escape fraction (f$_c$=0.1). Dashed curves illustrate our sensitivity limits 
(5$\sigma$) to the surface density of star-forming sources using our 
present analyzed data (top curve) and that for the completed survey 
(bottom curve). A lower limit to the abundance of star-forming galaxies 
required to build the $z\simeq$6 massive galaxies described in \cite{Eg05} ,\cite{Ey05}, 
\cite{HY05} over the period  7$<$z$<$20 is also  displayed. Taken together 
with Fig. 1 these massive systems implies a very early period of intense 
activity sufficient for cosmic reionization.}
\label{fig:fluxcal}
\end{figure}

\section{An HST+Spitzer Search for Strongly-Lensed z=7-10 Galaxies}

To complement the Ly$\alpha$ search, we have recently begun a HST large 
program (110 orbits) that   
will survey six cluster fields totaling 10~arcmin$^2$
in three wide-band filters (ACS-WFC/F850LP, NIC3/F110W, NIC3/F160W) to
characterize the faint, star-forming galaxies responsible for the
completion of cosmic reionization at $z>7$.  We achieve the
necessary depth by boosting the sensitivity of HST with
magnification from powerful lensing clusters at intermediate
redshifts (Figure 6). 

\begin{figure}[!t]
\centerline{\psfig{file=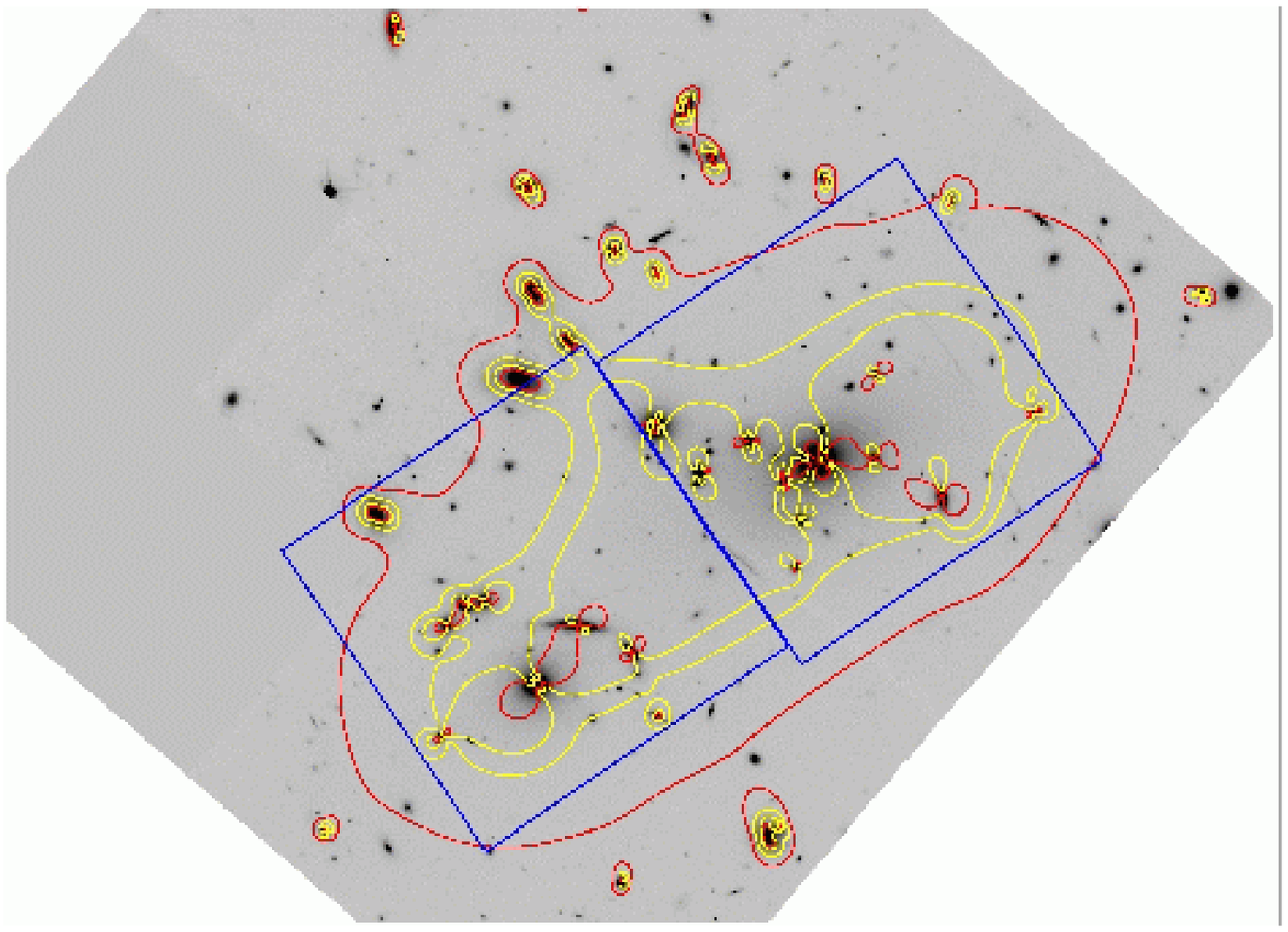,width=5.4in}}
\caption{Deployment of HST NICMOS images (squares) for Abell 2219
-  one of the target clusters. The outer curve defines the area
within which, as a result of
lensing magnification, the proposed survey will go deeper than
the NICMOS UDF campaign. The regions inside the yellow
curves represent areas of extreme ($\times$ 30) magnification.
}
\label{fig:fluxcal}
\end{figure}

We will select $z>7$ sources using well-established color-selection
techniques based on the neutral hydrogen break at rest-frame 
Lyman~$\alpha$ (e.g.,\cite{B04a}, \cite{B04b}; \cite{St04}).  
Incorporating the lensing magnification, more than half our survey area
will reach unlensed star-formation rates of $<$1M$_\odot$~yr$^{-1}$ 
(for nominal source assumptions), i.e. better than the NICMOS 
UDF survey \cite{B04b}. Although lensing reduces the surveyed 
volume by the linear magnification, our total volume at UDF depths is still 
almost 25\%  of the NICMOS UDF field.  Moreover, 2.5~arcmin$^2$ is imaged
more than one magnitude deeper in many sightlines and the fidelity of our 
sample will be assured by the well-understood geometrical effects of 
lensing.

Both observational evidence \cite{YW04} and theoretical 
conjecture \cite{BL02} suggest the counts of faint $z>7$ 
galaxies should be steeply rising at the UDF limit. Clearly at $z\simeq7$ 
only the most luminous galaxies are presently detectable \cite{B04b}.
 Extrapolating one magnitude deeper (Fig.~7) 
suggests we will reach a surface density of 11~arcmin$^2$ and find at least 5 
sources fainter than the UDF limit, approaching the minimal luminosity
density necessary for reionization.

\begin{figure}
\centerline{\psfig{file=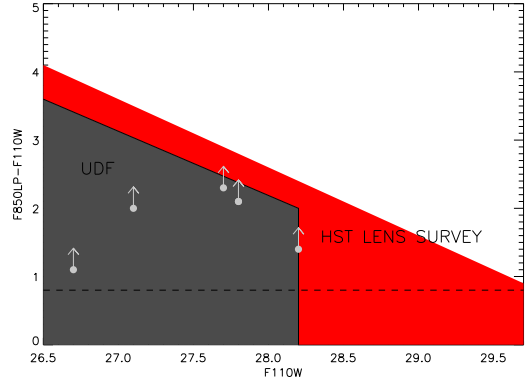,width=3in}}
\caption{Comparison of proposed sensitivity limits for the 
HST NICMOS/Spitzer deep survey (assuming a typical
lens magnification of 30) with those of \cite{B04b} in the
UDF (grey region).   The z$\sim$7 sources identified by \cite{B04b}
are plotted as light grey limits.  We will therefore explore new parameter
space that was inaccessible to the UDF.  Specifically, we will achieve
more stringent limits on F850LP-F110W colors and be sensitive to sources
down to $F110W{\simeq}30$.
}
\label{fig:cmag}
\end{figure}

A major bonus of our program is that magnified sources are brought 
within reach of the Spitzer Space Telescope. All of our target clusters have 
been imaged by IRAC at 3.6 and 4.5~$\mu$m, to the same 5-$\sigma$ 
point-source sensitivity of 1~$\mu$Jy as Abell 2218. The extended 
wavelength baseline offered by the combination of HST and Spitzer 
can uniquely characterize physical properties (age, star formation 
history) for z$\simeq$7 objects that would be too faint to otherwise 
be studied  \cite{Eg05}. By analogy with our analysis in 
Abell 2218, we can verify whether long-lived sources are the norm 
at $z\simeq$6-7. Tantalizingly, the combination of IRAC and NICMOS 
photometry also offers the prospect of searching for their more 
distant ancestors, z$\simeq$8-10,  via dropouts selected at the 
F110W/F160W boundary.

\end{document}